\documentclass[dvipdfm,epsf]{iopart}
\usepackage{iopams,amsfonts,rotating,graphicx,fancybox,fancyhdr,bm,eucal}

\eqnobysec

\begin{document}

\title[A note on the Pfaffian integration theorem]
        {A note on the Pfaffian integration theorem}

\author{Alexei Borodin$^1$ and Eugene Kanzieper$^2$}

\address{
    $^1$ Department of Mathematics, California Institute of Technology, CA 91125, USA\\
    $^2$ Department of Applied Mathematics, H.I.T. -- Holon Institute of
    Technology, Holon 58102,
    Israel} \eads{\mailto{borodin@caltech.edu} and
    \mailto{eugene.kanzieper@weizmann.ac.il}}

\begin{abstract}
Two alternative, fairly compact proofs are presented of the Pfaffian
integration theorem that is surfaced in the recent studies of
spectral properties of Ginibre's Orthogonal Ensemble. The first
proof is based on a concept of the Fredholm Pfaffian; the second
proof is purely linear algebraic. \newline\newline PACS numbers:
02.10.Yn, 02.50.-r, 05.40.-a\newline
doi:~10.1088/1751-8113/40/36/F01
\end{abstract}

\section{Introduction}
In the recent studies of spectral properties of Ginibre's Orthogonal
Ensemble (Ginibre 1965) of real asymmetric random matrices
\footnote[1]{For a discussion of physical applications of Ginibre's
random matrices, the reader is referred to the detailed paper by
Akemann and Kanzieper (2007).}, the following theorem was presented
by Kanzieper and Akemann (2005) and Akemann and Kanzieper (2007):

{\bf Pfaffian Integration Theorem.} {\it Let $\pi({\rm d}z)$ be an
arbitrary measure on $z\in {\mathbb C}$ with finite moments. Define
the function $Q_n(z,w) = {\boldsymbol {\underline q}}(z)
\,{\boldsymbol {\mu}}\, {\boldsymbol {\underline q}}^{\rm T}(w)$ in
terms of the vector ${\boldsymbol {\underline q}}(z) =
(q_0(z),\cdots, q_{n-1}(z))$ composed of arbitrary polynomials
$q_j(z)$ of the $j$-th order, and of an $n \times n$ antisymmetric
matrix ${\boldsymbol {\mu}}$. Then
\begin{eqnarray}\fl
    \label{theorem-a}
    \frac 1{\ell!}\prod_{j=1}^\ell \, \int_{ {\mathbb C}} \pi({\rm d}z_j)\,\,
    {\rm pf\,} \left[
    \begin{array}{cc}
      Q_{n}(z_i, z_j) & Q_{n}(z_i, {\bar z}_j) \\
      Q_{n}({\bar z}_i, z_j) & Q_{n}({\bar z}_i, {\bar z}_j) \\
    \end{array}
    \right]_{1\le i,j \le \ell}
    =
    e_\ell\left(
    \frac{1}{2} {\rm tr}(
    {\boldsymbol {\upsilon}}^1),\cdots,\frac{1}{2} {\rm tr}
    ({\boldsymbol {\upsilon}}^\ell)
    \right), \nonumber
\end{eqnarray}
where $e_\ell(p_1,\dots,p_\ell)$ are the elementary symmetric
functions \footnote[2]{Up to a factorial prefactor, the elementary
symmetric functions $e_\ell(p_1,\cdots, p_\ell)$ coincide with the
zonal polynomials $Z_{(1^\ell)}(p_1,\cdots, p_\ell)=\ell!
\,e_\ell(p_1,\cdots, p_\ell)$ appearing in the original formulation
of the theorem.} written as polynomials of the power sums, and
$n\times n$ matrix ${\boldsymbol {\upsilon}}$ is ${\boldsymbol
{\upsilon}} = {\boldsymbol {\mu}}{\boldsymbol {g}}$ with
\begin{eqnarray}
{\boldsymbol {g}} = \int_{\mathbb C} \pi({\rm d}z)
\,
({\boldsymbol {\underline q}}^{\rm T}(\bar z)  {\boldsymbol {\underline q}}(z)
-
{\boldsymbol {\underline q}}^{\rm T}(z)  {\boldsymbol {\underline q}}(\bar{z})). \nonumber
\end{eqnarray}
}\newline This theorem has been a key ingredient of the recent
calculation (Kanzieper and Akemann (2005) and Akemann and Kanzieper
(2007)) of the probability $p_{n,k}$ to find exactly $k$ real
eigenvalues in the spectra of $n\times n$ real asymmetric random
matrices drawn from Ginibre's Orthogonal Ensemble. An earlier
attempt to address the same problem is due to Edelman
(1997).\newline
\newline{\bf Remark 1.1.} The explicit form of
$e_\ell(p_1,\dots,p_\ell)$ is well known (see, e.g., Macdonald 1998):
\begin{eqnarray}
\label{SymPol}
     e_\ell(p_1,\cdots,p_\ell) =
    (-1)^\ell   \sum_{|\blambda|=\ell} \, \prod_{j=1}^g
    \frac{1}{\sigma_j!}
\left(
    - \frac{p_{\ell_j}}{\ell_j}
\right)^{\sigma_j}.
\end{eqnarray}
The notation
$\blambda = (\ell_1^{\sigma_1},\cdots,
    \ell_g^{\sigma_g})$ stands for the frequency representation of
    the partition $\blambda$ of the size $|\blambda|=\ell$. It implies
    that the part $\ell_j$ appears
    $\sigma_j$ times so that $\ell = \sum_{j=1}^g \ell_j\, \sigma_j$,
    where $g$ is the number of nonzero parts of
the partition.

An immediate corollary of Eq. (\ref{SymPol}) is the identity
\begin{eqnarray}
\label{zgf}
    \sum_{\ell=0}^\infty {\tau^\ell}\, e_\ell(p_1,\cdots, p_\ell) =
    \exp \left(
        \sum_{j=1}^\infty (-1)^{j-1} \tau^j\,\frac{ p_j}{j}
    \right).
\end{eqnarray}
\newline\newline
{\bf Remark 1.2.} The Pfaffian integration theorem can be viewed as
a generalisation of the Dyson integration theorem (Dyson 1970,
Mahoux and Mehta 1991) for the case where the quaternion kernel
$\boldsymbol{ {\mathcal Q}}_n(z,w)$ represented by the $2\times 2$
matrix \footnote[3]{See Appendix for the notation.}
\begin{eqnarray}
\label{mq}
    \Theta[\boldsymbol{ {\mathcal Q}}_n(z,w)] =
    {\boldsymbol{\tilde J}}^{-1}
     \left(
    \begin{array}{cc}
      Q_{n}(z, w) & Q_{n}(z, {\bar w}) \\
      Q_{n}({\bar z}, w) & Q_{n}({\bar z}, {\bar w}) \\
    \end{array}
    \right) \nonumber
\end{eqnarray}
does {\it not} satisfy the projection property
\begin{eqnarray}
    \int_{{\mathbb C}} {\rm d}\pi(w)\, \boldsymbol{ {\mathcal Q}}_n(z,w) \, \boldsymbol{ {\mathcal Q}}_n(w,z^\prime)
    = \boldsymbol{ {\mathcal Q}}_n(z,z^\prime). \nonumber
\end{eqnarray}

The original proof (Akemann and Kanzieper, 2007) of the theorem
involved an intricate topological interpretation of the ordered
Pfaffian expansion combined with the term-by-term integration that
spanned dozens of pages. In the present contribution, we provide two
alternative, concise proofs of slight variations of the Pfaffian
integration theorem. They are formulated in the form of the two
theorems and represent the main result of our note.
\newline\newline\newpage
{\bf Theorem 1.} {\it Let $(X,m)$ be a measure space, and the
vectors ${\boldsymbol {{\underline \varphi}^+}}(x)$ and ${\boldsymbol
{{\underline \varphi}^-}}(x)$ be composed of measurable functions
${\boldsymbol {{\underline \varphi}^\pm}}(x)=(\varphi^{\pm}_0(x), \cdots,
\varphi^{\pm}_{n-1}(x))$ from $X$ to ${\mathbb C}$. Define the
functions
\begin{eqnarray}
    \Phi_n^{\pm \pm}(x,y) = {\boldsymbol {{\underline \varphi}^\pm}}(x)\, {\boldsymbol {{\mu}}}\,
     {\boldsymbol {{\underline \varphi}}}^{\pm\;{\rm T}}(y), \nonumber
\end{eqnarray}
where ${\boldsymbol {{\mu}}}$ is an $n \times n$ antisymmetric
matrix. Then \footnote[4]{In what follows, we assume that our
measures are such that all integrals are finite.},
\begin{eqnarray}\fl
 \frac 1{\ell!}\prod_{j=1}^\ell \int_{X} m({\rm d}x_j)\,\,
    {\rm pf\,} \left[
    \begin{array}{cc}
      \Phi_n^{++}(x_i, x_j) & \Phi_n^{+-}(x_i, x_j) \\
      \Phi_n^{-+}(x_i, x_j) & \Phi_n^{--}(x_i, x_j) \\
    \end{array}
    \right]_{1\le i,j \le \ell} \nonumber \\
     \label{thm1}
    \qquad\qquad\qquad\qquad=
    e_{\ell}\left(
    \frac{1}{2} {\rm tr}(
    {\boldsymbol {\upsilon}}^1),\cdots,\frac{1}{2} {\rm tr}
    ({\boldsymbol {\upsilon}}^\ell)
    \right),
\end{eqnarray}
where the $n\times n$ matrix ${\boldsymbol {\upsilon}}$ is
${\boldsymbol {\upsilon}} = {\boldsymbol {\mu}}{\boldsymbol {g}}$ with
\begin{eqnarray}
\label{g-matrix}
{\boldsymbol {g}} = \int_{X} m({\rm d}x)
\,
({\boldsymbol {\underline \varphi}^{-\; {\rm T}}}(x) {\boldsymbol {\underline \varphi}^+}(x)
-
{\boldsymbol {\underline \varphi}}^{+\; \rm T}(x) {\boldsymbol {\underline \varphi}}^{-}(x)). 
\end{eqnarray}
}
\newline
{\bf Theorem 2.} {\it In the notation of Theorem 1, assume that the
matrix ${\boldsymbol {{\mu}}}$ is invertible (hence, $n$ is even).
Then
\begin{eqnarray}\fl
\label{t2}
\sum_{\ell=0}^{n/2} \frac{\tau^\ell}{\ell!}\prod_{j=1}^\ell \int_{X}
m({\rm d}x_j)\,\,
    {\rm pf\,} \left[
    \begin{array}{cc}
      \Phi_n^{++}(x_i, x_j) & \Phi_n^{+-}(x_i, x_j) \\
      \Phi_n^{-+}(x_i, x_j) & \Phi_n^{--}(x_i, x_j) \\
    \end{array}
    \right]_{1\le i,j \le \ell} \nonumber \\
    \qquad\qquad\qquad\qquad\qquad\qquad
= {\rm pf\,}{\boldsymbol {\mu}}\cdot{\rm pf\,}[\,{\boldsymbol
{\mu}}^{-1\;{\rm T}}-\tau \boldsymbol {g}].
\end{eqnarray}
}\newline
{\bf Remark 1.3.} The equivalence of Theorem 1 and Theorem 2 is easily established. Indeed,
multiplying both sides of Eq.~(\ref{thm1}) by $\tau^\ell$ and summing up
over $\ell$ from 0 to $\infty$ with the help of Eq.~(\ref{zgf}), one finds that the
right--hand side turns into
$$
\exp\left(\frac{1}{2}\sum_{j=1}^\infty (-1)^{j-1} \tau^j\frac{{\rm tr} (\boldsymbol
{\upsilon}^j)}{j} \right)=\sqrt{\det
(\boldsymbol{I}+\tau\boldsymbol {\upsilon})}={\rm pf\,}{\boldsymbol
{\mu}}\cdot{\rm pf\,}[\,{\boldsymbol {\mu}}^{-1\,{\rm T}}-\tau \boldsymbol
{g}],
$$
given that $\boldsymbol{\mu}$ is invertible. This proves the
equivalence for invertible (nondegenerate) $\boldsymbol{\mu}$.
Theorem~1 with degenerate $\boldsymbol{\mu}$ of even size follows by
the limit transition, and decreasing the size $n$ by $1$ is achieved
by setting $\varphi_{n-1}^{\pm}\equiv 0$ and nullifying the last,
$n$-th, row and column in the matrix ${\boldsymbol \mu}$.

\section{Fredholm Pfaffian Proof of the Theorem 1.}

For $\ell \in {\mathbb Z}^+$, define the infinite sequence
\begin{eqnarray} \fl
\label{pi-L}
    {\sigma}_\ell (n)=
     \prod_{j=1}^\ell \int_{X} m({\rm d}x_j)\,\,
    {\rm pf\,} \left[
    \begin{array}{cc}
      \Phi_n^{++}(x_i, x_j) & \Phi_n^{+-}(x_i, x_j) \\
      \Phi_n^{-+}(x_i, x_j) & \Phi_n^{--}(x_i, x_j) \\
    \end{array}
    \right]_{1\le i,j \le \ell},
\end{eqnarray}
supplemented by $\sigma_0(n)=1$, and consider the series
\begin{eqnarray}
    {\mathcal S}(\tau; n) = \sum_{\ell=0}^\infty \frac{\tau^\ell}{\ell!}\, \sigma_\ell(n).
\end{eqnarray}
By definition [Eq.~(\ref{fp})] introduced by Rains (2000), the
function ${\mathcal S}(\tau;n)$ is the Fredholm Pfaffian on the
measure space $(X,m)$
\begin{eqnarray}
\label{R-lambda}
    {\mathcal S}(\tau;n) = {\rm pf}_X\big[ \boldsymbol{J}+\tau \boldsymbol{ \Phi}_n \big]
    = \sqrt{{\rm det}_X \big[ \boldsymbol{I}  - \tau \boldsymbol{ J} \boldsymbol{ \Phi}_n\big]}.
\end{eqnarray}
(See Appendix for the matrix notation used.) Here,
$\boldsymbol{\Phi}_n$ is the $2\times 2$ matrix kernel
\begin{eqnarray}
    \boldsymbol{\Phi}_n(x,y) =
        \left(
    \begin{array}{cc}
      \Phi_n^{++}(x, y) & \Phi_n^{+-}(x, y) \\
      \Phi_n^{-+}(x, y) & \Phi_n^{--}(x, y) \\
    \end{array}
    \right)
\end{eqnarray}
that can also be written as
\begin{eqnarray}
            \boldsymbol{\Phi}_n(x,y) &=&
        \left(
    \begin{array}{cc}
      \boldsymbol{\underline \varphi}^+(x) \boldsymbol{\underline \psi}^{+\; {\rm T}}(y) &
      \boldsymbol{\underline \varphi}^+(x) \boldsymbol{\underline \psi}^{-\; {\rm T}}(y) \\
      \boldsymbol{\underline \varphi}^-(x) \boldsymbol{\underline \psi}^{+\; {\rm T}}(y) &
      \boldsymbol{\underline \varphi}^-(x) \boldsymbol{\underline \psi}^{-\; {\rm T}}(y)
      \\
    \end{array}
    \right) \nonumber \\
    &=&
            \left(
              \begin{array}{c}
                \boldsymbol{\underline \varphi}^+(x) \\
                \boldsymbol{\underline \varphi}^-(x) \\
              \end{array}
            \right)\,
            \left(
              \begin{array}{cc}
                \boldsymbol{\underline \psi}^{+\; {\rm T}}(y), & \boldsymbol{\underline \psi}^{-\; {\rm T}}(y) \\
              \end{array}
            \right),
\end{eqnarray}
where the vector $\boldsymbol{\underline \psi}^{\pm}(x)$ is
$\boldsymbol{\underline \psi}^{\pm}(x) = \boldsymbol{\underline
\varphi}^{\pm}(x)\, \boldsymbol{\mu}^{\rm T}$.

The matrix $\boldsymbol{ I}  - \tau \boldsymbol{J}
\boldsymbol{\Phi}_n$ appearing under the sign of the Fredholm
determinant can be represented as $\boldsymbol{ I} + \boldsymbol
{A} \boldsymbol{ B}$ with
\begin{eqnarray}
    \boldsymbol{A}(x,\alpha) &=& - \tau\,{\boldsymbol J}         \left(
              \begin{array}{c}
                \boldsymbol{\underline \varphi}^+(x) \\
                \boldsymbol{\underline \varphi}^-(x) \\
              \end{array}
            \right) = \tau
       \left(
              \begin{array}{c}
                -\boldsymbol{\underline \varphi}^-(x) \\
                \boldsymbol{\underline \varphi}^+(x) \\
              \end{array}
            \right),
             \\
    \boldsymbol{B}(\alpha,y) &=&
    \left(
              \begin{array}{cc}
                \boldsymbol{\underline \psi}^{+\; {\rm T}}(y), & \boldsymbol{\underline \psi}^{-\; {\rm T}}(y) \\
              \end{array}
            \right).
\end{eqnarray}
Following Tracy and Widom (1998), one observes the ``needlessly
fancy'' general relation ${\rm det}[\boldsymbol{I} +
\boldsymbol{AB}] = {\rm det}[\boldsymbol{I} + \boldsymbol{BA}]$ that
holds for arbitrary Hilbert-Schmidt operators $\boldsymbol{A}$ and
$\boldsymbol{B}$. They may act between different spaces as long as
the products make sense. In the present context, ${\rm
det}[\boldsymbol{I} + \boldsymbol{AB}]$ is the Fredholm determinant
${\rm det}_X[\boldsymbol{I} + \boldsymbol{AB}]$ whilst the
determinant ${\rm det}[\boldsymbol{I} + \boldsymbol{BA}]$ is that of
the $n\times n$ matrix
\begin{eqnarray}
    {\boldsymbol I} + \tau \int_X m({\rm d}x)
    (
                \boldsymbol{\underline \psi}^{-\; {\rm T}}(x) \boldsymbol{\underline \varphi}^+(x)
                -
                \boldsymbol{\underline \psi}^{+\; {\rm T}}(x) \boldsymbol{\underline \varphi}^-(x)
    ) = {\boldsymbol I} + \tau {\boldsymbol \upsilon},
\end{eqnarray}
where $\boldsymbol{\upsilon}=\boldsymbol{\mu\, g}$ with ${\boldsymbol g}$ defined by Eq. (\ref{g-matrix}).

The above calculation allows us to write down the Fredholm Pfaffian
${\mathcal S}(\tau;n)$ in the form
\begin{eqnarray} \fl
    {\mathcal S}(\tau;n) = \sqrt{{\det }\left[{\boldsymbol I} + \tau {\boldsymbol \upsilon}
        \right]}
     &=& \exp\left[ \frac{1}{2}\,{\rm tr}\log \left(\boldsymbol{I}+ \tau {\boldsymbol {\upsilon}}\right) \right]
     \nonumber \\
     &=& \exp\left[ \frac{1}{2}
        \sum_{j=1}^{\infty} (-1)^{j-1} \tau^j \frac{{\rm tr}
        ({\boldsymbol { \upsilon}}^j) }{j}
     \right].
\end{eqnarray}
Identity Eq.~(\ref{zgf}) concludes the proof. $\qquad\Box$

\section{Linear Algebraic Proof of the Theorem 2.}

A linear algebraic proof of Theorem 2 is based on the works by
Ishikawa and Wakayama (1995, 2000) and de Bruijn (1955) as
formulated in Sections \ref{P-1} and \ref{P-2}. Section \ref{P-3}
contains a proof of Theorem 2.

\subsection{Minor summation formulae and identities by Ishikawa and Wakayama}
\label{P-1}

Let ${\boldsymbol T}$ be any $M \times N$ matrix, and let $[m]$
denote the set $\{1,2,\cdots,m\}$ for a positive integer $m \in
{\mathbb Z}^+$. For $n$-element subsets $I=\{i_1 < \cdots < i_n\}
\subseteq [M]$ and $J=\{j_1 < \cdots < j_n\} \subseteq [N]$ of row
and column indices, let ${\boldsymbol T}^I_J={\boldsymbol
T}^{i_1\cdots i_n}_{j_1\cdots j_n}$ denote the submatrix of ${\boldsymbol T}$
obtained by picking up the rows and columns indexed by $I$ and $J$.
In this notation, the following three lemmas hold \footnote[5]{Lemma
2 is a reformulation of Theorem 3.1 by Ishikawa and Wakayama
(2000)}.

{\bf Lemma 1 (Ishikawa and Wakayama 1995).} {\it Let $M\le N$ and
assume $M$ is even. For any $M\times N$ matrix ${\boldsymbol A}$ and
any $N\times N$ antisymmetric matrix ${\boldsymbol B}$, one has
\begin{eqnarray}
\sum_{I\subseteq [N],\,\# I =M}
{\rm pf\,} [{\boldsymbol B}_I^I] \det [{\boldsymbol A}_I^I] = {\rm
pf\,} [\,\boldsymbol{ A B A}^{\rm T}].
\end{eqnarray}}

{\bf Lemma 2 (Ishikawa and Wakayama 2000).} {\it Let
$\boldsymbol{A}$ be an $N\times N$ invertible antisymmetric matrix.
Then, for any $I \subseteq [N]$, one has
\begin{eqnarray}
{\rm pf\,} [{\boldsymbol A}]\,{\rm pf\,}[({\boldsymbol A}^{-1})^I_I] = (-1)^{|I|} {\rm pf\,}
[{\boldsymbol A}^{\overline{I}}_{\overline{I}}],
\end{eqnarray}
where $\overline{I}\subseteq [N]$ stands for the complementary of
$I$, and $|I|$ denotes the sum of the elements of $I$,
$|I|=\sum_{i\in I}i$.}

{\bf Lemma 3 (Ishikawa and Wakayama 2000).} {\it Let ${\boldsymbol
A}$ and ${\boldsymbol B}$ be $N\times N$ antisymmetric matrices.
Then
\begin{eqnarray}
{\rm pf\,}[{\boldsymbol A}+{\boldsymbol B}] = \sum_{r=0}^{\lfloor N/2 \rfloor} \sum_{I \subseteq [N], \; \# I = 2r}
(-1)^{|I|-r} {\rm pf\,}[{\boldsymbol A}_I^{I}] \, {\rm pf\,}[\boldsymbol{B}_{\overline{I}}^{\overline{I}}].
\end{eqnarray}
Here $\lfloor n \rfloor$ denotes the integer part of $n$. }

In what follows, we will need the following corollary.

{\bf Corollary 1.} {\it Let $\boldsymbol{A}$ and $\boldsymbol{B}$ be
$N\times N$ antisymmetric matrices, and $\boldsymbol{A}$ is
invertible. Then
\begin{eqnarray}
\sum_{I} {\rm pf\,}[\boldsymbol{A}_{I}^{I}] \,
 {\rm pf\,}[\boldsymbol{B}_{I}^{I}] = {\rm pf\,}[\boldsymbol{A}]\cdot {\rm pf\,}[\boldsymbol{A}^{-1\, {\rm T}}
 +\boldsymbol{B}].
\end{eqnarray}}

{\it Proof.} Using Lemma 3 and Lemma 2 (in this order), we write
down:
\begin{eqnarray} 
{\rm pf\,}[{\boldsymbol A}^{-1\, {\rm T}}+\boldsymbol{B}] = \sum_{r=0}^{\lfloor N/2 \rfloor} \sum_{I \subseteq [N], \; \# I = 2r}
(-1)^{|I|} {\rm pf\,}[(\boldsymbol{A}^{-1})_I^{I}] \, {\rm pf\,}[\boldsymbol{B}_{\overline{I}}^{\overline{I}}] \nonumber \\
= \frac{1}{{\rm pf\,}[\boldsymbol{A}]}\sum_{r=0}^{\lfloor N/2 \rfloor} \sum_{I \subseteq [N], \; \# I = 2r}
 {\rm pf\,}[\boldsymbol{A}_{\overline{I}}^{\overline{I}}] \, {\rm pf\,}[\boldsymbol{B}_{\overline{I}}^{\overline{I}}] \nonumber \\
 = \frac{1}{{\rm pf\,}[{\boldsymbol A}]}\sum_{I} {\rm pf\,}[\boldsymbol{A}_{I}^{I}] \,
 {\rm pf\,}[\boldsymbol{B}_{I}^{I}].
\end{eqnarray}
This concludes the proof.  $\qquad\Box$

\subsection{de Bruijn integration formula}
\label{P-2}

{\bf Lemma 4 (de Bruijn 1955).\;}{\it Let $(X,m)$ be a measure
space, and the vectors ${\boldsymbol {{\underline \varphi}^+}}(x)$
and ${\boldsymbol {{\underline \varphi}^-}}(x)$ be composed of
measurable functions ${\boldsymbol {{\underline
\varphi}^\pm}}(x)=(\varphi^{\pm}_0(x), \cdots,
\varphi^{\pm}_{2\ell-1}(x))$ from $X$ to ${\mathbb C}$. Then
\begin{eqnarray*}
\prod_{j=1}^\ell \int_X m({\rm d}x_j)\,
\det
\left[
\begin{array}{c}
    \varphi_j^{+}(x_i) \\
    \varphi_j^{-}(x_i)
  \end{array}
\right]_{0 \le i \le 2\ell-1,\,1\le j\le \ell}
=\ell!\, {\rm pf}\left[{\boldsymbol g}^{\rm T}\right],
\end{eqnarray*}
where the $2\ell\times 2\ell$ matrix ${\boldsymbol {g}}$ is
\footnote[8]{Compare to Eq. (\ref{g-matrix}).}
\begin{eqnarray}
{\boldsymbol {g}} = \int_{X} m({\rm d}x)
\,
({\boldsymbol {\underline \varphi}^{-\; {\rm T}}}(x) {\boldsymbol {\underline \varphi}^+}(x)
-
{\boldsymbol {\underline \varphi}}^{+\; \rm T}(x) {\boldsymbol {\underline \varphi}}^{-}(x)). \nonumber
\end{eqnarray}
}

\subsection{Proof of Theorem 2.}
\label{P-3}

In the notation of Theorem 1, let us set $\boldsymbol{A}$ to be the
$2\ell\times n$ matrix
\begin{eqnarray}
    \boldsymbol{A} = \left(
  \begin{array}{c}
    \varphi_j^{+}(x_i) \\
    \varphi_j^{-}(x_i)
  \end{array}
\right)_{1\le i \le \ell, 0 \le j \le n -1}
\end{eqnarray}
and identify $\boldsymbol{B}$ with the $n\times n$ matrix
${\boldsymbol \mu}$,
\begin{eqnarray}
{\boldsymbol B}={\boldsymbol \mu}.
\end{eqnarray}
Noting that
\begin{eqnarray}
    {\rm pf\, }[\boldsymbol{ABA}^{\rm T}]= {\rm pf\,} \left[
    \begin{array}{cc}
      \Phi_n^{++}(x_i, x_j) & \Phi_n^{+-}(x_i, x_j) \\
      \Phi_n^{-+}(x_i, x_j) & \Phi_n^{--}(x_i, x_j) \\
    \end{array}
    \right]_{1\le i,j \le \ell},
\end{eqnarray}
we make use of the l.h.s. of Lemma 1, to write down the expansion
\begin{eqnarray} \fl
        {\rm pf\,} \left[
    \begin{array}{cc}
      \Phi_n^{++}(x_i, x_j) & \Phi_n^{+-}(x_i, x_j) \\
      \Phi_n^{-+}(x_i, x_j) & \Phi_n^{--}(x_i, x_j) \\
    \end{array}
    \right]_{1\le i,j \le \ell} \nonumber \\
    =
    \sum_{I \subseteq [n],\, \# I = 2\ell}
    {\rm pf\,}[{\boldsymbol \mu}_I^I]
    \, {\rm det}
    \left[
        \begin{array}{c}
    \varphi_j^{+}(x_i) \\
    \varphi_j^{-}(x_i)
  \end{array}
    \right]_{1\le i\le \ell,\, j\in I}.
\end{eqnarray}
Consequently, the l.h.s. of Eq. (\ref{t2}) reduces to (note that $n$
is even by the hypothesis)
\begin{eqnarray} \fl
\label{aux-step}
    \sum_{\ell=0}^{n/2} \frac{\tau^\ell}{\ell!}\,
    \sum_{I \subseteq [n],\, \# I = 2\ell}
    {\rm pf\,}[{\boldsymbol \mu}_I^I]
    \,
    \prod_{i=1}^\ell \int_{X} m({\rm d}x_i)\,
    {\rm det}
    \left[
        \begin{array}{c}
    \varphi_j^{+}(x_i) \\
    \varphi_j^{-}(x_i)
  \end{array}
    \right]_{1\le i\le \ell,\, j\in I} \nonumber \\
    =
    \sum_{\ell=0}^{n/2} \tau^\ell
    \sum_{I \subseteq [n],\, \# I = 2\ell}
    {\rm pf\,}[{\boldsymbol \mu}_I^I]
    \,
    {\rm pf\,}[({\boldsymbol g}^{\rm T})_I^I].
\end{eqnarray}
Here, we have used Lemma 3. By Corollary 1, the r.h.s. of Eq. (\ref{aux-step}) is equivalent to
\begin{eqnarray}
    {\rm pf\,}[{\boldsymbol \mu}]\cdot
    {\rm pf\,}[{\boldsymbol \mu}^{-1\, {\rm T}} + \tau {\boldsymbol g}^{\rm T}]
    =
    {\rm pf\,}[{\boldsymbol \mu}]\cdot
    {\rm pf\,}[{\boldsymbol \mu}^{-1\, {\rm T}} - \tau {\boldsymbol g}].
\end{eqnarray}
This concludes the proof.    $\qquad\Box$

\section*{Acknowledgements}
EK is grateful to G. Akemann, a joint work with whom on the
integrable structure of Ginibre's Orthogonal Ensemble has triggered
this study. AB was partially supported by the NSF grants DMS-0402047
and DMS-0707163. EK acknowledges a partial support by the Israel
Science Foundation through the grant No 286/04.
\smallskip

\appendix
\section*{Appendix. The Fredholm Pfaffian}  \label{apa}
\renewcommand{\theequation}{A.\arabic{equation}}
\addcontentsline{toc}{section}{Appendix. Fredholm Pfaffian}

Let ${\boldsymbol K}(x,y)$ be a $2\times 2$ matrix kernel
\begin{eqnarray}
   {\boldsymbol K}(x,y) = \left(
               \begin{array}{cc}
                 K_{11}(x,y) & K_{12}(x,y) \\
                 K_{21}(x,y) & K_{22}(x,y) \\
               \end{array}
             \right)
\end{eqnarray}
which is antisymmetric under the change of its arguments,
${\boldsymbol K}(x,y)=-\boldsymbol{K}^{\rm T}(y,x)$, and yet another
$2\times 2$ matrix kernel ${\boldsymbol J}(x,y)$ be
\begin{eqnarray}
    {\boldsymbol J}(x,y) = I(x,y)\, {\boldsymbol{\tilde J}}, \;\;\; I(x,y) = \delta_{x,y},\;\;\;
    {\boldsymbol{\tilde J}} =  \left(
                            \begin{array}{cc}
                              0 & 1 \\
                              -1 & 0 \\
                            \end{array}
                          \right).
\end{eqnarray}

(i) The {\it Fredholm Pfaffian} ${\rm pf}_X[{\boldsymbol
J}+\boldsymbol{K}]$ on the measure space $(X,m)$ is defined via the
series (Rains 2000)
\begin{eqnarray} \label{fp}
    {\rm pf}_X[{\boldsymbol J}+\boldsymbol{K}] = \sum_{\ell=0}^\infty \frac{1}{\ell!}
    \prod_{j=1}^\ell
    \int_{X} m({\rm d}x_j) \, {\rm pf\,}[{\boldsymbol K}(x_i,x_j)]_{1\le i,j\le \ell}.
\end{eqnarray}

(ii) A more familiar {\it Fredholm determinant} ${\rm det}_X[ I+K ]$
of a scalar kernel $K$,
\begin{eqnarray}
    {\rm det}_X[I+ K] = \sum_{\ell=0}^\infty \frac{1}{\ell!}
    \prod_{j=1}^\ell \int_{X} m({\rm d}x_j)\, {\rm det}[K(x_i,x_j)]_{1\le i,j\le \ell},
\end{eqnarray}
appears to be a particular case of the Fredholm Pfaffian since
\begin{eqnarray}
    {\rm pf}_X \left[
        {\boldsymbol J} + \left(
              \begin{array}{cc}
                \varepsilon & K \\
                -K & 0 \\
              \end{array}
            \right)
    \right] = {\rm det}_X[ I+K].
\end{eqnarray}
Here, $\varepsilon$ is any antisymmetric scalar kernel.

(iii) The connection between Fredholm Pfaffian and Fredholm
determinant is given by
\begin{eqnarray}
\label{fpd}
    {\rm pf}_X[{\boldsymbol J}+{\boldsymbol K}]^2 = {\rm det}_X\left[ {\boldsymbol I}
    -{\boldsymbol J}{\boldsymbol K}\right],
\end{eqnarray}
where ${\boldsymbol I}$ is the $2\times 2$ matrix kernel
\begin{eqnarray}
    {\boldsymbol I}(x,y) = I(x,y)\, {\boldsymbol{\tilde I}}, \;\;\;
    {\boldsymbol{\tilde I}} =  \left(
                            \begin{array}{cc}
                              1 & 0 \\
                              0 & 1 \\
                            \end{array}
                          \right).
\end{eqnarray}

\section*{References}
\fancyhead{} \fancyhead[RE,LO]{References}
\fancyhead[LE,RO]{\thepage}

\begin{harvard}

\item[] Akemann G and Kanzieper E 2007
        Integrable structure of Ginibre's ensemble of real random matrices and a Pfaffian integration theorem
        {\it arXiv:~math-ph/0703019 (to appear in J. Stat. Phys.)}

\item[]  de Bruijn N G 1955
          On some multiple integrals involving determinants
          {\it J. Indian Math. Soc.} {\bf 19} 133

\item[] Dyson F J 1970
        Correlations between eigenvalues of a random matrix
        {\it Commun. Math. Phys.} {\bf 19} 235

\item[] Edelman A 1997
        The probability that a random real Gaussian matrix has $k$ real eigenvalues,
        related distributions, and the circular law
        {\it J. Multivar. Anal.} {\bf 60} 203

\item[] Ginibre J 1965
        Statistical ensembles of complex, quaternion, and real
        matrices
        \JMP {\bf 19} 133

\item[] Ishikawa M and Wakayama M 1995
        Minor summation formula of Pfaffians
        {\it Linear and Multilinear Algebra} {\bf 39} 285

\item[] Ishikawa M and Wakayama M 2000
        Minor summation formulas of Pfaffians, survey and a new identity
        {\it Adv. Stud. Pure Math.}
        {\bf 28} 133

\item[] Kanzieper E and Akemann G 2005
        Statistics of real eigenvalues in Ginibre's ensemble of random real
        matrices
        \PRL {\bf 95} 230201

\item[] Macdonald I G 1998
        {\it Symmetric Functions and Hall Polynomials} (Oxford: Oxford
        University Press)

\item[] Mahoux~G and Mehta~M~L 1991
        A method of integration over matrix variables
        {\it J. Phys. I (France)} {\bf 1} 1093

\item[] Rains~E~M 2000
        Correlation functions for symmetrized increasing subsequences
        {\it arXiv:~math.CO/0006097}

\item[] Tracy C A and Widom H 1998 Correlation functions, cluster functions and spacing distributions for random matrices
        {\it J. Stat. Phys.} {\bf 92} 809

\smallskip
\end{harvard}

\end{document}